\documentclass[reprint, prl, superscriptaddress, floatfix, noeprint]{revtex4-1}

\usepackage{hyperref}
\usepackage{graphicx}
\usepackage{amsmath}
\usepackage{wasysym}
\usepackage{amsfonts}
\usepackage{color}
\usepackage{verbatim}
\usepackage{upgreek}
\usepackage{bm}
\usepackage{textcomp}
\usepackage[symbol*]{footmisc}
\usepackage[final]{changes}
\usepackage{float}
\usepackage{braket}

\DefineFNsymbolsTM{otherfnsymbols}{%
  \textdagger    \dagger
  \textdaggerdbl \ddagger
  \textsection   \mathsection
}%

\setfnsymbol{otherfnsymbols}

\newcommand{\beq}{\begin{equation}}
\newcommand{\eeq}{\end{equation}}
\newcommand{\bea}{\begin{eqnarray}}
\newcommand{\eea}{\end{eqnarray}}
\newcommand{\bal}{\begin{align}}
\newcommand{\eal}{\end{align}}

\newcommand{\fig}[1]{Fig.\,\ref{#1}}

\newcommand{\WS}{WS$_2$}
\newcommand{\Wf}{W$_{\mathrm{4f}}$}
\newcommand{\G}{$\mathrm{\Gamma}$}
\newcommand{\GdW}{G$\Delta$W}
\newcommand{\dW}{$\Delta$W}
\newcommand{\eeff}{\ensuremath{\varepsilon_{\mathrm{eff}}}}

\definecolor{ablue}{rgb}{0.1,0.3,0.65}

\definecolor{blue}{rgb}{0,0.15,0.75}

\begin{document}
\title{Rigid band shifts in two-dimensional semiconductors through environmental screening}
\author{Lutz Waldecker$^*$} 
\email{waldecker@stanford.edu}
\affiliation{Dept. of Applied Physics, Stanford University,  348 Via Pueblo Mall, Stanford, California 94305, USA}
\affiliation{SLAC National Accelerator Laboratory, Menlo Park, California 94025, USA}
\author{Archana Raja$^*$} 
\email{archana.raja@berkeley.edu}
\affiliation{Dept. of Applied Physics, Stanford University,  348 Via Pueblo Mall, Stanford, California 94305, USA}\affiliation{SLAC National Accelerator Laboratory, Menlo Park, California 94025, USA}
\affiliation{Kavli Energy NanoScience Institute, University of California Berkeley, Berkeley, California 94720, USA}
\author{Malte R\"{o}sner$^*$} 
\affiliation{Institute for Molecules and Materials, Radboud University, 6525 AJ Nijmengen, The Netherlands}
\author{Christina Steinke} 
\affiliation{Institute for Theoretical Physics, University of Bremen, Otto-Hahn-Allee 1, 28359 Bremen, Germany}
\affiliation{Bremen Center for Computational Material Sciences, University of Bremen, Am Fallturm 1a, 28359 Bremen, Germany}
\author{Aaron Bostwick} 
\author{Roland J. Koch} 
\author{Chris Jozwiak} 
\affiliation{Advanced Light Source, E. O. Lawrence Berkeley National Laboratory, Berkeley, California 94720, USA}
\author{Takashi Taniguchi}
\author{Kenji Watanabe}
\affiliation{National Institute for Materials Science, Tsukuba, Ibaraki 305-004, Japan}
\author{Eli Rotenberg} 
\affiliation{Advanced Light Source, E. O. Lawrence Berkeley National Laboratory, Berkeley, California 94720, USA}
\author{Tim O. Wehling} 
\affiliation{Institute for Theoretical Physics, University of Bremen, Otto-Hahn-Allee 1, 28359 Bremen, Germany}
\affiliation{Bremen Center for Computational Material Sciences, University of Bremen, Am Fallturm 1a, 28359 Bremen, Germany}
\author{Tony F. Heinz} 
\email{theinz@stanford.edu}
\affiliation{Dept. of Applied Physics, Stanford University,  348 Via Pueblo Mall, Stanford, California 94305, USA}\affiliation{SLAC National Accelerator Laboratory, Menlo Park, California 94025, USA}

\begin{abstract}

We investigate the effects of environmental dielectric screening on the electronic dispersion and the band gap in the atomically-thin, quasi two-dimensional (2D) semiconductor \WS\ using correlative angle-resolved photoemission and optical spectroscopies, along with first-principles calculations. 
We find the main effect of increased environmental screening to be a reduction of the band gap, with little change to the electronic dispersion of the band structure. 
These essentially rigid shifts of the bands results from the special spatial structure of the changes in the Coulomb potential induced by the dielectric environment in the 2D limit. 
Our results suggest dielectric engineering as a non-invasive method of tailoring the band structure of 2D semiconductors and provide guidance for understanding the electronic properties of 2D materials embedded in multilayer heterostructures.

\end{abstract}

\maketitle
\date{\today}


In monolayers of atomically-thin, quasi two-dimensional (2D) semiconductors, the intrinsic screening of Coulomb interactions is reduced compared to their bulk crystals, since electric field lines between charges extend significantly outside the material. 
As a result, exciton binding energies are enhanced, reaching values of several hundreds of meV in the transition metal dichalcogenides (TMDCs) \cite{Ramasubramaniam2012, Berkelbach2013, Qiu2013, He2014, Chernikov2014, Ye2014, Wang2018}. 
For the same reason, materials in close proximity to the monolayers enhance the effective screening of charge carrier interactions. 
By embedding atomically-thin materials in different dielectric environments, their band gaps, as well as exciton binding energies, can therefore be modified on an energy scale of the exciton binding energies themselves \cite{Rosner2016,Stier2016b,Raja2017}.
This sensitivity becomes particularly important in vertical heterostructures of 2D materials and enables a non-invasive way of designing nanoscale functionality, such as lateral heterojunctions, through the spatial control of substrate dielectrics \cite{Rosner2016,Raja2017,Utama2019}.

To exploit the full potential of tailoring Coulomb interactions through control of the dielectric environment, it is critical to understand its impact not only on the band gap but also on the valence and conduction band dispersions.
The dispersion determines such basic properties as the effective masses of the carriers and the energy differences between different valleys within the Brillouin zone, and also affects the relative alignment between the valence and conduction bands of a homogeneous monolayer with spatially varying external dielectric screening.
To date, experimental studies of dielectric engineering have mainly focused on optical spectroscopy or electronic transport measurements of TMDC monolayers, which only probe a small fraction of the full Brillouin zone. 
In general, however, perturbations to a material do not have the same effect on electronic states of different orbital character and can be expected to modify the band structure in different parts of the Brillouin zone differently.

\begin{figure*}[!bht]
    \begin{center}
        \includegraphics[width=\textwidth]{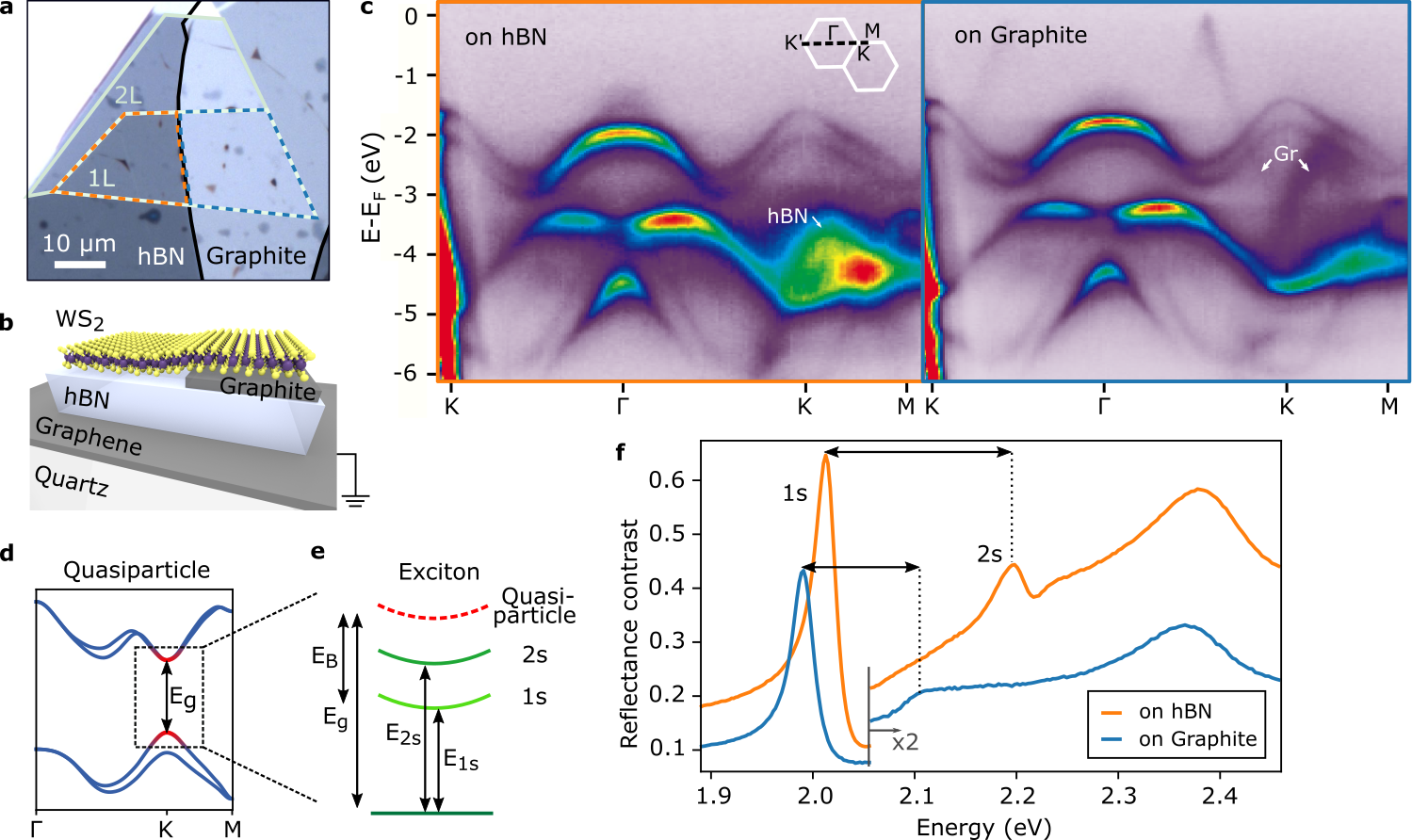}
        \caption{a) Optical micrograph of monolayer \WS\ (dotted white line) straddling distinct dielectric environments on hBN (orange) and graphite (blue) b) Schematic of the sample geometry on a transparent conductive substrate, enabling both optical spectroscopy and ARPES measurements. c) Photoemission intensity maps along the K'-\G-K-M direction (see inset) of monolayer \WS\ on hBN and on graphite. $\pi$-bands of hBN and back-folded graphite bands are visible, no signatures of hybridization with \WS\ bands are observed. d) Sketch of the band structure, showing the direct quasiparticle band gap at the K(K') points. The quasiparticle states around the K point that form the ground state A exciton transition are highlighted in red. e) Schematic of exciton ground and excited state transitions, showing the relationship between exciton transition energies, exciton binding energy $\mathrm{E_B}$ and quasiparticle band gap $\mathrm{E_g}$ corresponding to the K(K') point transition. f) Room temperature reflectance contrast spectrum of monolayer \WS\ on hBN (orange) and on graphite (blue).  }
        \label{fig:1}
    \end{center}
\end{figure*}
Here, through a combination of experiment and theory, we provide a generalized picture of the consequences of dielectric screening for the band structure of 2D semiconductors. 
By combining angle-resolved photoemission spectroscopy with micrometer spatial resolution (\textmu-ARPES) and optical spectroscopy of the exciton states of monolayer \WS\ on different substrates, we find that the predominant effect of external screening is a band gap renormalization through a rigid shift of the occupied and unoccupied bands relative to each other.
These rigid shifts are a result of the spatial structure of the changes in the Coulomb potential induced by the dielectric environment, which we elucidate with the aid of \emph{ab initio} \GdW\ calculations.  
Our results illustrate how non-local screening in 2D materials can yield a solid state analogue of molecular level renormalization in different solvent environments or on surfaces \cite{Neaton2006,Thygesen2009}.


Monolayers of the semiconducting TMDC \WS\ were exfoliated from bulk crystals and transferred such that they partially cover two different substrates, hexagonal Boron Nitride (hBN) and graphite (for experimental and sample fabrication details, see Supplementarty Information).
An optical micrograph of a typical sample and a schematic of the sample geometry used in the experiments are shown in Figs.\,\ref{fig:1}\,a and b. 
 
\begin{figure*}[tbh]
    \begin{center}
        \includegraphics[width=\textwidth]{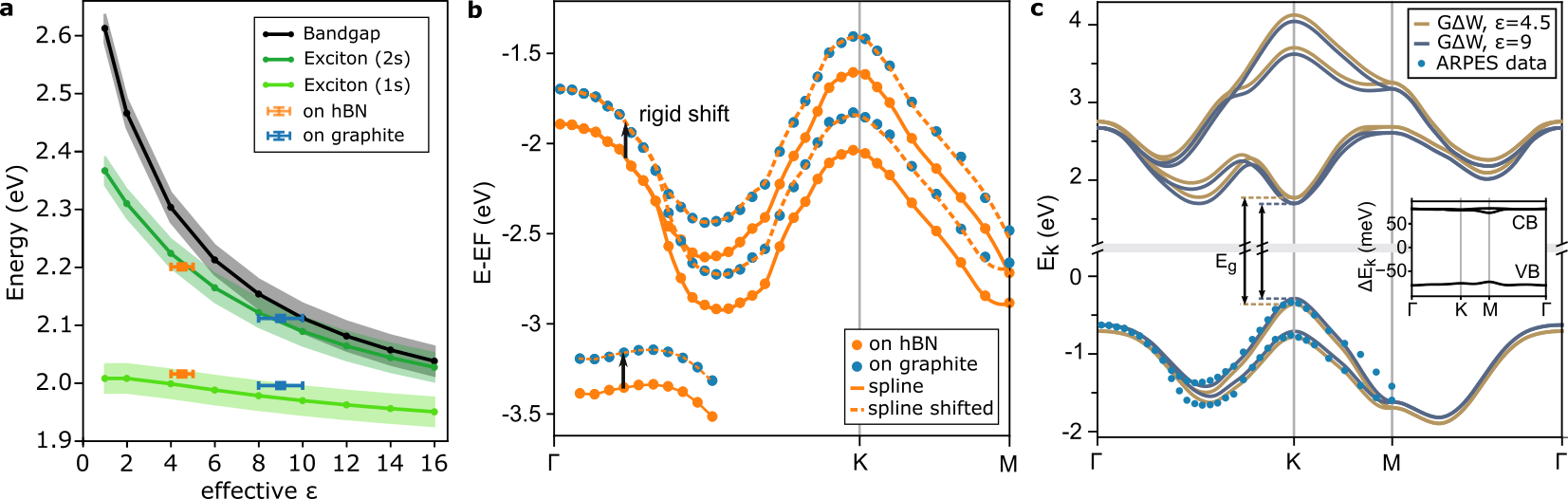}
        \caption{a) Calculated quasiparticle band gap and 1s and 2s exciton energies as a function of effective dielectric constant \eeff\ (lines) and experimental values of the exciton energies on hBN and graphite (markers). b) Experimental valence band positions on hBN (orange points) and graphite (blue points) along the $\Gamma$-K-M direction. A spline through the data points on hBN is shown in solid orange. This spline can be rigidly shifted by ca. $0.18\,$eV to overlay the data points on graphite (dashed orange line). c) Calculated band structure of the highest valence and lowest conduction band (lines) and experimental results (circles). The experimental data has been aligned to the calculated curves at the $\Gamma$ points. Inset: Difference of the dispersion between the conduction and valence bands of \WS\ supported on substrates with $\eeff=4.5$ and $\eeff=9$.}
        \label{fig:2}
    \end{center}
\end{figure*}

Using \textmu-ARPES, we measure the valence band dispersion and the separation to shallow core levels, including \Wf.
Two examples of room temperature photoemission intensity maps of the valence bands of \WS\ in the K'-$\Gamma$-K-M direction of \WS\ are shown in \fig{fig:1}\,c. 
Signatures of the respective substrates appear in both spectra, such as the $\pi$-band of hBN and replicas of graphite bands extending to the Fermi energy E$_F$.
However, no signs of hybridization between \WS\ and the respective substrate bands are observed.

By measuring the exciton states on the same samples, we obtain information on quasiparticle band gaps at the K points (\fig{fig:1}\,d).
Typical reflectance contrast spectra, approximately proportional to the monolayer absorption, are shown in \fig{fig:1}\,f. 
As the oscillator strength in 2D semiconductors mainly resides in their excitonic absorption features, a series of prominent peaks is seen in the spectra. 
We identify the two lowest lying features as the 1s and 2s exciton states of the A exciton transition \cite{Chernikov2014}. 
While the quasiparticle band gap is not directly accessible, it is proportional to the 1s-2s separation $\Delta_{12}$ (\fig{fig:1}\,e) \cite{Chernikov2014, Raja2017, Cho2018}.
The exciton binding energy and the quasiparticle band gap are sensitive to dielectric screening from the immediate environment of hBN and graphite, which is reflected in both the shifting of the exciton peaks and, more importantly, the reduction of the 1s-2s separation \cite{Raja2017}. 
This is also evidence of close contact between \WS\ and the hBN/graphite substrates. 
The lack of charged exciton signatures and narrow linewidths in the optical spectra indicate  low doping levels \cite{Mak2013}, consistent with the ARPES data in which the Fermi level is observed within the band gap.

We now analyze the experimentally measured dispersion and the band gap renormalization in conjunction with first-principles calculations.
We calculate the band structure of a freestanding \WS\ monolayer in the GW approximation, from which the band gap in vacuum ($\varepsilon = 1$) is obtained (see SI for details) in good agreement with Ref.~\cite{haastrup_computational_2018}. 
The change in the band gap is then calculated using a combination of the Wannier function continuum electrostatics (WFCE) \cite{rosner_wannier_2015} and \GdW\ approaches \cite{Rohlfing2010, Winther2017}, where \dW\ is the environmentally-induced change to the Coulomb potential resulting from a dielectric substrate described by an effective dielectric constant \eeff.
in this way, we reduce the complex dielectric function of the substrates to an effective screening constant \eeff, independent of momentum and frequency. 

The calculated change of the quasiparticle band gap, i.e. the difference between the valence band maximum and the conduction-band minimum at the K-points, is shown in \fig{fig:2}\,a.
To compare this result to the measured exciton positions, we additionally solve the Wannier equation for the screened potentials and obtain the binding energies of 1s and 2s exciton states (see SI). 
We find good agreement between experimental and calculated exciton positions for $\eeff \approx 4.5$ and $\eeff \approx 9$ for hBN and graphite respectively (see \fig{fig:2}~c).
These values are in reasonable agreement with previously reported values \cite{Andersen2015}. 
We note that \eeff\ of the same substrate can be different for other 2D semiconductors. 
The calculated band gap renormalization upon changing the dielectric substrate from $\epsilon_{\mathrm{eff}}=4.5$ to $\epsilon_{\mathrm{eff}}=9$ is found to be 140\,meV. 
We conclude that the band gap of monolayer \WS\ on hBN compared to graphite is approximately 140\,meV larger.

To elucidate how the environmental screening affects the electronic dispersion, we determine the band positions from the ARPES data in \fig{fig:1}\,c by fitting energy distribution curves (EDCs) of the valence bands at each recorded parallel momentum $k_x$ and accounting for detector distortions (see SI).
Intriguingly, a spline through the data points on hBN can be rigidly shifted to overlay the data points on graphite within the experimental error of approximately 25\,meV. 
In particular, the relative alignment of the K-points with respect to $\Gamma$ is determined as 280(280)$ \pm 10$\,meV on hBN (graphite) and the spin-orbit splitting at the K-points \cite{Zhu2011} is 440(430)$ \pm 20$\,meV.
The effective masses in the valence bands are determined from quadratic fits as 2.45(2.55)$\pm 0.05$ $m_e$ at $\Gamma$, 0.48(0.48)$\pm0.05$ $m_e$ in the upper and 0.64(0.78)$\pm0.1$ $m_e$ in the lower valence band at the K points on hBN (graphite).

The calculated valence- and conduction-band dispersions for the two values of \eeff\ are shown in \fig{fig:2}\,c together with the the experimental data points on graphite for comparison.
Since the absolute band energies in photoemission can be influenced by external fields, we align the experimental bands to the GW calculations at their $\Gamma$-point energies.
The calculated curves closely follow the measured dispersion, with small deviations roughly halfway between $\Gamma$ and K as well as close to M.
These discrepancies may arise from the difficulty of fitting two bands where their separation is small, along with approximations used in the calculations. 
Since deviations occur in a region of strong orbital hybridization, we expect it to be particularly sensitive to small errors in lattice relaxation.

From our calculations of the dispersion, it is clear that the main effect of the environmental screening is a rigid shift of occupied and unoccupied bands, as also observed in Ref. \cite{Winther2017}.
The shift is symmetric in the valence and conduction bands, which is to first order intrinsic to the 2D slab geometry as discussed in detail later.
The change in dispersion is visualized in the inset of \fig{fig:2}\,c, where the difference between the band energies $\Delta E_k = E_{k,\varepsilon=4.5}-E_{k,\varepsilon=9}$ is plotted. 
In the calculations, the deviations from a rigid shift are smaller than 5 meV across the Brillouin zone, which is less than 5\% of the band gap renormalization and is consistent with our experimental observations. 

\begin{figure*}[bth]
    \begin{center}
        \includegraphics[width=\textwidth]{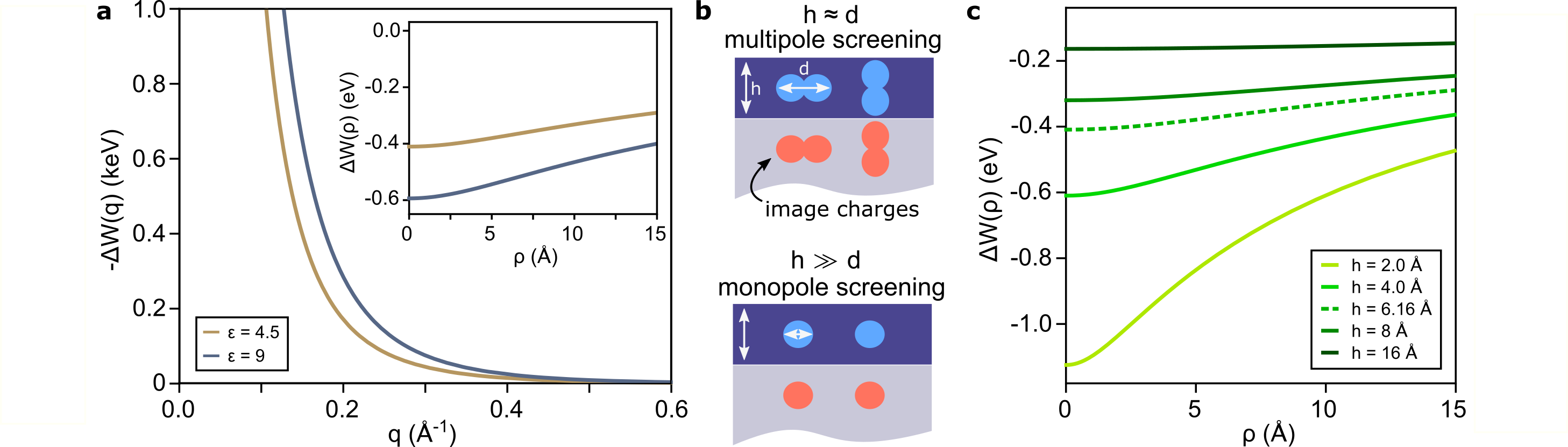}
        \caption{a) Environmentally-induced change in Coulomb potential $\Delta W$ in \WS\. The change is peaked at zero momentum, corresponding to an almost flat change in real space (inset). b) Sketch of multi- vs. monopole screening due to image charges in the substrate for different layer heights $h$ to orbital extension $d$ ratios. For $h\gg d$, monopole screening dominates and bands are shifted rigidly. c) Change of the Coulomb potential for different layer thicknesses $h$ and $\eeff = 4.5$. For small $h$, the change shows a pronounced spatial dependence, which can cause asymmetric band shifts.}
        \label{fig:3}
    \end{center}
\end{figure*}


These essentially rigid shifts can be understood from the change of the Coulomb interaction profile $\Delta W_\varepsilon(\rho) = W_\varepsilon(\rho) - W_{V}(\rho)$ as resulting from the screening environment with dielectric constant $\varepsilon$ with respect to the interaction of the freestanding layer in vacuum (V) and its effects to the \GdW\ band structure (see SI). In \fig{fig:3}\,a we show examples for $\Delta W_{\varepsilon = 4.5}$ and $\Delta W_{\varepsilon = 9}$ as functions of the real-space coordinate $\rho$ (inset) and the momentum-transfer $q$. Since $W_{V}$ is always larger than $W_\varepsilon$, $\Delta W_\varepsilon$ is by definition negative. In real space we find nearly constant potential profiles for $\rho < 10\,$\AA\, which approaches zero for larger $\rho$. This behavior of $\Delta W_\varepsilon(\rho)$ results from the non-local screening properties of the 2D slab with finite height $h$. In the case of purely local screening we expect $\Delta W_\varepsilon^{loc}(\rho) \propto \frac{1}{\varepsilon \rho} - \frac{1}{\rho} = \frac{1-\varepsilon}{\varepsilon \rho}$ to diverge for small $\rho$. Here, however, the two dielectric interfaces (top and bottom side of the WS$_2$ layer) are separated by $\pm h/2$ from the center of the slab and create an alternating infinite series of image charges localized at distances $h\gtrsim 6\,$\AA. The corresponding contributions to $\Delta W_\varepsilon(\rho)$ are of the form $\propto\frac{1}{\sqrt{h^2+\rho^2}}$ \cite{Cho2018} and thus flat at small $\rho$. Therefore, it is necessary to fully take the effective height $h$ into account.
The well-known approximation of the Keldysh potential $W(\rho) \propto \alpha^{-1}[ H_0(\rho/\alpha) - J_0(\rho/\alpha)]$ \cite{cudazzo_dielectric_2011}, which is only valid for $\rho \gg h$, is therefore not capable of describing this  particular change in interaction profile.

The flat interaction profile $\Delta W_\varepsilon(\rho)$ in real space translates to a strongly peaked profile in momentum space being zero for $q \gtrsim 0.4\, \text{\AA}^{-1}$. For the following analysis we can thus approximate $\Delta W_\varepsilon(\rho) \approx \Delta W_\varepsilon(\rho=0) = \gamma_\varepsilon$ and $\Delta W_\varepsilon(q) \approx \gamma_\varepsilon \delta(q)$. Importantly, this type of interaction does not distinguish between different orbital characters and cannot cause any inter-band scattering.

In this case, the electronic self-energy $\Sigma_{G\Delta W}$, which describes the changes in the electronic quasiparticle dispersions in the WS$_2$ layer due to changes in the environmental screening, greatly simplifies in the \GdW\ approximation. For electrons in band $\lambda$ with momentum $k$ it reads
\begin{align}
	\Sigma^{\lambda}_{G\Delta W}(k,\omega) 
		= &\frac{i}{2\pi}
		   \int dq \int d\omega' \frac{\Delta W_\varepsilon(q, \omega')}{\omega + \omega' + i\delta - E_{k-q}^\lambda}. \label{eqn:SigmaGdWG}		         
\end{align}
In the static Coulomb-hole plus screened-exchange (COHSEX) approximation, this self-energy $\Sigma_{G\Delta W}$ can be split into two terms resulting from poles in $G$ and in $\Delta W$ yielding $\Sigma^{\lambda}_{SEX}(k) \approx - \gamma_\varepsilon  n_F(E_k^\lambda)$ and $\Sigma^{\lambda}_{COH}(k) \approx \frac{\gamma_\varepsilon}{2}$, respectively, where $n_F$ is the Fermi function. The SEX part shifts only occupied states up in energy and the COH terms shifts all bands down by $\gamma_\varepsilon/2$. Importantly, these self-energies are independent of $k$ for completely filled (empty) valence (conduction) bands, since the Fermi functions depend only on band index $\lambda$ but not on $k$ here. The quasiparticle dispersions under the influence of environmental screening then read
\begin{align}
	E_{k,\varepsilon}^\lambda = E_{k,\varepsilon=1}^\lambda + \gamma_\varepsilon \left[n_F(E_{k,\varepsilon=1}^\lambda) - \frac{1}{2}\right].
\end{align}
The bands shift as a whole with no changes to the dispersion and the band gap is symmetrically reduced by $\gamma_\varepsilon$ equally for all momenta $k$. 
Experimentally, we observe these rigid shifts down to the core-levels \Wf\ (see SI) and also find them in our full COHSEX calculations [using the full orbital-dependent $\Delta W^{\alpha\beta}(q)$] presented in \fig{fig:2}. 

In order to change the band shapes or to induce asymmetric band shifts, significant deviations from the approximation $\Delta W_\varepsilon(q) \approx \gamma_\varepsilon \delta(q)$ are needed. 
Thus, either $\gamma_\varepsilon$ must become orbital-dependent, or $\Delta W_\varepsilon(q) \propto \delta(q)$ must break down. 
This is controlled by the ratio between the effective WS$_2$ height $h$ and the orbital extension $d$. For small $h \ll d$, for example, the multipole-pole screening by the image charges differentiates between different orbital characters (see \fig{fig:3}\,b), and $\gamma_\varepsilon^{\alpha\beta}$ becomes orbital-dependent. 
Also, for small $h$ the change to the Coulomb potential $\Delta W_\varepsilon(\rho)$ starts to show a spatial structure (see \fig{fig:3}\,c), and $\Delta W_\varepsilon(q) \propto \delta(q)$ becomes inaccurate. 
Thus, by reducing the effective height or increasing the orbital extension, non-rigid-shift modifications may occur. 
In the case of the TMDCs, the transition metal d orbitals are ``shielded" by the surrounding chalcogen atoms which increases the effective height and reduces non-rigid-shift effects. 
In effectively thinner materials with multi-orbital band-edge characters these effects could, however, be stronger.

Thus the validity of the approximation $\Delta W_\varepsilon(q) \approx \gamma_\varepsilon \delta(q)$ is an intrinsic property of the monolayer and certainly holds for \WS. 
We therefore do not expect any deviations from the rigid-shift-like changes to the band structure, even if the environmental screening shows a significant frequency dependence like in the case of graphite \cite{lin_plasmons_1997}. 
As shown above, there are indeed no additional changes to the valence bands for graphite (Fig.\ref{fig:2}\,b) so that our static theory adequately interpolates the experimental dispersion (Fig.\ref{fig:2}\,c).


In conclusion, we have built a lateral heterojunction by exposing a homogeneous \WS\ monolayer to spatially-separated dielectric environments. 
With the help of our combined experimental-theoretical studies we are able to show that the main change of the electronic properties of \WS\ between both environments is a band gap opening and a rigid shift of its valence and conduction bands.
The non-local nature of the screening leads to almost constant changes of the Coulomb potential in \WS, which translate to a symmetric opening and closing of the band gap.
This mechanism is consistent with recently reported data from transport measurements across a similar dielectrically-engineered lateral heterojunction \cite{Utama2019}.
The observed rigid shifts are in stark contrast to other methods of band gap engineering, such as ion-doping, in which the upper valence band at K is modified and the spin-splitting of the bands increases \cite{Katoch2017} or the application of strain, which results in a change of the energetic alignment of different valleys \cite{Yun2012, Conley2013}.
Our results establish dielectric engineering as a non-invasive way of modifying the quasiparticle band gap of 2D semiconductors and will help understand phenomena on the interfaces between 2D semiconductors and other materials.

\nocite{perdew_generalized_1996} 
\nocite{blochl_projector_1994}
\nocite{kresse_efficiency_1996,kresse_efficient_1996}
\nocite{shishkin_implementation_2006}
\nocite{mostofi_wannier90:_2008}
\nocite{liu_three-band_2013,steinhoff_influence_2014}
\nocite{rosner_wannier_2015}
\nocite{haastrup_computational_2018}
\nocite{schuler_optimal_2013}

\section{Acknowledgements}
We would like to thank Simone Latini for helpful hints for solving the 2D Schr\"{o}dinger equation. The spectroscopic studies were supported by the US Department of Energy, Office of Science, Basic Sciences, Materials Sciences and Engineering Division, under Contract DE-AC02-76SF00515 for analysis and by the Gordon and Betty Moore Foundation’s EPiQS Initiative through Grant No. GBMF4545. L.W. and M.R. acknowledge support by the Alexander von Humboldt Foundation. A.R. gratefully acknowledges funding through the Heising-Simons Junior Fellowship within the Kavli Energy NanoScience Institute at the University of California, Berkeley. T.W. and C.S. acknowledge support by the European Graphene Flagship and DFG via GRK 2247. K.W. and T.T. acknowledge support from the Elemental Strategy Initiative conducted by the MEXT, Japan, A3 Foresight by JSPS and the CREST (JPMJCR15F3), JST. This research used resources of the Advanced Light Source, which is a DOE Office of Science User Facility under contract no. DE-AC02-05CH11231.


%

\end{document}